**Challenges in materials and devices for Resistive-Switching-based Neuromorphic Computing**

Javier del Valle[1], Juan Gabriel Ramírez[2], Marcelo J. Rozenberg[3] and Ivan K. Schuller[1]

[1] *Department of Physics and Center for Advanced Nanoscience, University of California-San Diego, La Jolla, California 92093, USA*
[2] *Department of Physics, Universidad de los Andes, Bogotá 111711, Colombia*
[3] *Laboratoire de Physique des Solides, CNRS-UMR 8502, Université Paris-Sud, Orsay 91405, France*

**Abstract**

This tutorial describes challenges and possible avenues for the implementation of the components of a solid-state system, which emulates a biological brain. The tutorial is devoted mostly to a charge-based (i.e. electric controlled) implementation using transition metal oxides materials, which exhibit unique properties that emulate key functionalities needed for this application. In the Introduction, we compare the main differences between a conventional computational machine, based on the Turing-von Neumann paradigm, to a Neuromorphic machine, which tries to emulate important functionalities of a biological brain. We also describe the main electrical properties of biological systems, which would be useful to implement in a charge-based system. In Chapter II, we describe the main components of a possible solid-state implementation. In Chapter III, we describe a variety of Resistive Switching phenomena, which may serve as the functional basis for the implementation of key devices for Neuromorphic computing. In Chapter IV we describe why transition metal oxides, are promising materials for future Neuromorphic machines. Theoretical models describing different resistive switching mechanisms are discussed in Chapter V while existing implementations are described in Chapter VI. Chapter VII presents applications to practical problems. We list in Chapter VIII important basic research challenges and open issues. We discuss issues related to specific implementations, novel materials, devices and phenomena.  The development of reliable, fault tolerant, energy efficient devices, their scaling and integration into a Neuromorphic computer may bring us closer to the development of a machine that rivals the brain.

**I. INTRODUCTION**

We are living in the information age. We have powerful computers that are interconnected to even more powerful databases that provide seemingly endless information and enable actions at a distance. This technological revolution is based on two pillars developed during the past century: hardware computers and software codes. The first one is supported by impressive technological progress in electronics, including transistors, microprocessors, memories, LEDs, etc. The second one is supported by equally impressive advances in computer science, including programming languages, communication protocols and cryptography, just to name a few. These two avenues of technological progress merged into the information superhighway that we use and continue to develop every day. "Moore's law", the doubling of transistors in dense circuits every two years, has fueled the explosion in the use and manipulation of data in everyday life. However, it is commonly agreed that in the next 15-25 years a "Moore's crisis" will develop, in which the continuous improvement in computational power and the decrease in cost will end.[1]

The vast majority of today's commercial computers follow the von Neumann architecture model. This digital electronic computer was described by John von Neumann in 1945 based on work from Eckert and Mauchly.[2] Independently Turing was developing logical and practical ideas on how to implement a Universal Computing machine.[3] This so-called Turing-von Neumann (TvN) paradigm features several units and their interconnections as shown in the figure 1. It is characterized as a store-program machine, as it keeps both the data and the program instructions in a common storage space. The main components are a Central Processing Unit (CPU), a Memory Unit connected to the outside world by input and output devices. They can be easily recognized in the familiar desktop machines today. The CPU is further composed of a control unit, an arithmetic and logic unit and registers. The CPU fetches data and programing instructions from the memory unit. These instructions and data are further acted upon by the control unit, and the operations are then executed by the arithmetic and logic unit. Input and output devices are the interfaces with the operator or the external world.

Some pioneer ideas regarding artificial intelligence (AI) go back to seminal thoughts of John von Neumann[4] who first pointed out that brains cognitive power does not emerge from accurate digital calculations but rather from a collective form of computation involving large number of slow, imprecise and unreliable analog components.[5,6] Our brains are amazingly more efficient at recognizing patterns than powerful digital computers. To distinguish the image of a cat from a tiger is a hard task for a computer, but not for a child. AI is making strides thanks to "neuromorphic" inspired concepts developed decades ago, such as neural networks[7] and convolution filters, using the tremendous calculational power of current conventional computers.[8] These AI systems are achieving remarkable feats, from beating the best chess and go players to recognize faces in a crowd.[9] However, the hardware requirements to run artificial neural networks are extremely high, both for calculation speed and power consumption, which represents a significant impediment for conventional computer technology. While the brain of a chess player uses of the order of 20 watts, a supercomputer requires a million times more to face the same challenge. Therefore, a different approach to computing, probably inspired in neuromorphic concepts, will undoubtedly lead to more energy efficient machines and an improvement in the global power consumption.[10,11]

The massive amount of information at our disposal poses new challenges. How to make sense of it? How to recognize patterns? How to make decisions from vast but often conflicting, incomplete or imprecise data?

These challenges open an opportunity for the emergence of a disruptive technology. Breaking away from the classical Turing-von Neumann machine paradigm to implement a new type of bio-inspired ("neuromorphic") electronic devices or cognitive hardware[6] may allow for the implementation of artificial neural networks. These new machines may be based on new types of devices, such as artificial synapses, axons, dendrites and neurons, to enable the construction of "neuromorphic" circuits with AI capabilities.

In a neuromorphic computer, contrary to a TvN machine, there is not a clear-cut separation between the unit executing the operations (calculations or logical) and the memory. A neuromorphic system is characterized by simple units with a high degree of interconnectivity (figure 1), imitating the brain. The units perform simple operations as compared to the arithmetic and logic unit of a TvN machine. These constituent units, which emulate "neurons" and "synapses", are interconnected in a so-called Neural Network.

Another key difference between TvN and neuromorphic machines is that the formers operate in eminently serial mode, while the latter are intrinsically parallel. This feature has many consequences. Perhaps the main one is that the operation of TvN machines require a high degree of precision. Since their operation is serial, the initial data undergoes a long sequence of successive manipulations, which are susceptible to error build-up. Hence, they often require calculations with 16 significant figures ("double precision") although typically the precision needed for the end result is only one part in a thousand. Neuromorphic systems operating in parallel have units that integrate the input of a rather large number of interconnected units. The statistical nature inherent in their operation makes these systems less vulnerable to accumulation of errors. In addition, the operation speed of a serial system is given by the speed of each basic operation times the number of operations. The first is quite fast in modern machines ($\sim 10^{-9}$s) but the former can be quite large in complex calculations. In contrast, parallel systems may have relatively slow operating units but a large number of them and significantly shorter calculations, which may allow for faster computations.

**Von-Neumann architecture**          **Neuromorphic architecture**

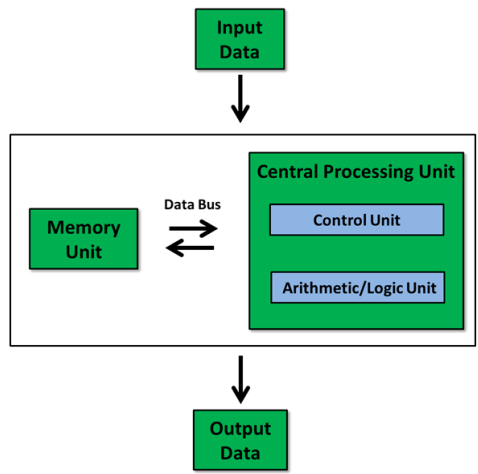 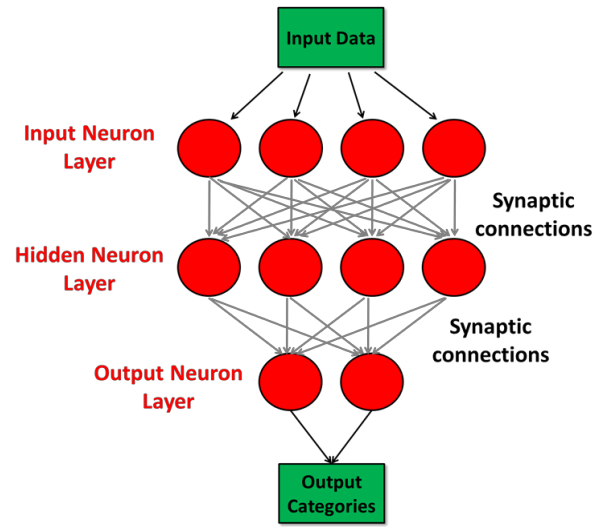

Figure 1. Comparison between von-Neumann and neural network computing architectures.

The purpose of a neuromorphic hardware[12] is to emulate a biological system by mimicking the most relevant electric functions of biological neuronal devices. The purpose is not to reproduce exactly a biological system but to extract essential features, which may allow for the development of a functioning machine. Thus, we will outline below the most important properties of the main components of a biological-neural system

***Neurons*** The central elements of biological neural circuits are the neurons,[13] active elements whose function is to produce electric signals (typically spikes) as a response to an excitation. This response is non-linear, which generates a spike only if the excitation voltage exceeds a threshold. The response is also time dependent: the neuron is more likely to fire if excitation signals arrive more frequently. This behavior is commonly referred to as "leaky, integrate and fire" (LIF). The membrane of neuron acts as a leaky capacitor whose voltage builds up as incoming pulses arrive, but slowly discharges (leaks) over time. If the threshold of the excitatory voltage is exceeded, the neuron activates and fires (figure 2).[14] In biological neurons the integration process is believed to occur within a region of the cell body called the "axon hillock".

***Axons and dendrites*** Neurons span out connections, like cables, called dendrites and axons (figure 2), which provide inter-neuronal connections and carry electric impulses from neuron to neuron. Dendrites are the receiving connections of the neuron, gathering all the input signals, while axons carry away the output generated by the neuron. Each neuron may have from one to hundreds of dendrites, and they may further branch into spines that can reach tens of thousands of connections. On the other hand, neurons solely have a single or maybe few axons. This results on a highly-interconnected network of neurons, through which the electric pulses generated by one neuron are sent to many others. It is important to point out the differences between axons and dendrites. Usually they are considered as mere transmission line equivalents, however, (axons in particular) can also implement information processing tasks.[12,15] Moreover, since synchronization between multiple neurons play a role in brain activity, long-range connectivity through axons may allow the coordinated behavior of remote neurons.

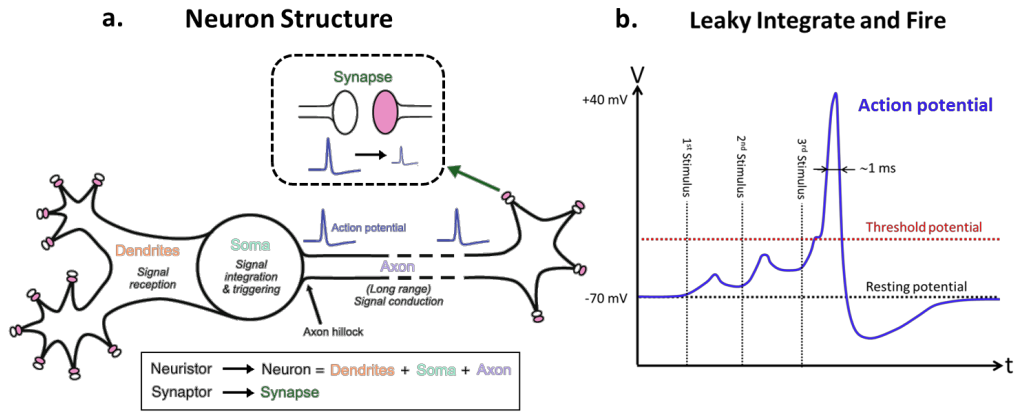

Figure 2: Schematic electrically relevant structure a) and voltage spiking, b) of a neuron showing the leaky-integrate and fire after a threshold potential is reached.

***Synapses*** Another key element of biological neuronal networks is the synapse, the connection point between axons and dendrites (figure 2). These synaptic connections regulate the fraction of the outgoing signal travelling through the axon which is transferred to the dendrite. In other words, they control how strongly interconnected two neurons are. This interconnection strength is known as "synaptic weight" and is how the memory of the network is stored. The weights also define the network's topology, as they control the propagation of signals across the synapses. These weights are not fixed but can change over time depending on their previous history. This property, known as "synaptic plasticity", provides the learning capability of the network. The initial study of this feature in biological systems is associated to Hebb, who synthetized it in the phrase "neurons that fire together, wire together." [16] The requirement for synaptic weights to modify their values according to the correlations in the electric spike activity of their connected neurons is called "spike-timing-dependent plasticity" (STDP).[14,17] Due to this mechanism, the synaptic weight of a given synapse is reinforced when the connected pre-neuron becomes active just before the post-neuron does. And conversely, the weight is decreased if the post-neuron becomes active before the pre-neuron. The first effect is called "long-term potentiation" (LTP) and the second one is "long-term depression" (LTD). [14]

The above discussion provides a starting point for the minimal functionalities expected from electronic neuromorphic devices. It seems clear that a bio-inspired system may be composed of four basic artificial devices emulating: neurons, synapses, axons and dendrites, wired with a high degree of inter-connectivity.

A very important issue is related to the energy consumption of a TvN machine as compared to a biological neural system. To develop a TvN machine that is able to perform similar tasks as the human brain, the number of devices must be increased by several orders of magnitude with respect to current ordinary computers. While it is not obvious how to compare the energy consumption of such a TvN machine with a biological brain, it is commonly accepted that the energy consumption of the former is considerable larger by many orders of magnitude. From the energetic point of view the increased number of devices poses two main challenges: i) the *local* energy consumption produces deleterious thermal interactions between the devices, and ii) the *global* energy consumption becomes prohibitive. Many estimates indicate that a few of these high-performance TvN machines would require a serious fraction of the world energy production. Thus, to develop ever more powerful machines, it is imperative to break away from the TvN paradigm.

Artificial Intelligence machines that attempt to emulate the computational behavior of neural systems are known as Neuromorphic Computers. Several implementations have been advanced and applied to the solutions of different computational problems. The most obvious one is to use conventional silicon based technology (CMOS) to implement selected neural functionalities.[18] Unfortunately, in order to obtain even the most rudimentary ones many classical CMOS devices are needed,[12] which greatly increases energy consumption. However, these kinds of systems, even in a reduced implementation, allow proving conceptual ideas of neuromorphic machines. Alternatively, completely new redesigns of computational machines have been developed in a variety of configurations. Spin torque oscillator-based machines have been developed recently, which rely on the temporal evolution of those devices and have been applied, for instance, to speech recognition.[19] Energy efficient hybrid superconducting-magnetic devices are being developed to simulate synapses.[20] And semiconductor-superconducting devices, using few photon light-emitting diodes connected with superconducting wires, have been proposed to emulate spiking neurons.[21] In this context, a particularly promising approach, that we shall describe below, attempts to emulate all-important neuro-biological functionalities in a single platform, which relies on the electrical response of certain carefully chosen compounds belonging to the family of quantum materials.[22]

In this review article, we shall discuss and describe how neuromorphic functionalities may be realized using Resistive Switching (RS), a physical phenomenon in which the resistance of a device depends on the applied voltage or/and its previous history.[23–25] Although this phenomenon has already been observed in many materials,[26,27] in this review we will focus on the promising transition metal oxides (TMO).[28–34] TMOs are attractive as they feature the two main types of resistive switching: volatile and non-volatile. In volatile switching, the resistance of the material changes when a high enough voltage or current is applied, returning to its original value afterwards. In non-volatile switching, the resistance of the material does not go back to its original value, but it is permanently modified. These two types of resistive switching allow implementing the basic hardware elements of neuromorphic systems that we shall denote: *neuristors* (which emulate the leaky, integrate and fire behavior of spiking neurons) and *synaptors* (which emulate the memory behavior of synapses).[35]

## II. DEVICES AND FUNCTIONALITIES

The terminology used in the field of Neuromorphic Computing is somewhat confusing. Because of this we will define the terms for the artificial ingredients of a charge based Neuromorphic Machine and the context in which we will use them as follows:

**Neuristors** are active elements of the network which emulate the computation functionality of neurons. They must feature a non-linear activation function, with an all or nothing response in the form of an electric spike. Volatile resistive switching enables this feature thanks to its threshold behavior: the material remains non-conductive until a threshold voltage (more precisely, an electric field) is surpassed and provokes the collapse of the resistance. Once the voltage is removed the device returns to its initial (non-conductive) resting state, perhaps with a controllable time constant that may be associated with the refractory period of neurons. A key functionality they must mimic is the Leaky, Integrate and Fire (LIF) behavior, which implies that the neuristor must keep some sort of short term memory that allows it to sum all inputs. [35,36] Such memory might come, for instance, from the relaxation time of the insulator-metal phase transition, as shown recently.[37] The neuristor must have one or several input and output transmission lines, corresponding to the dendrites and axons, respectively. These axons and dendrites could be in principle implemented by simple metallic interconnects, which seems to be the approach taken in most real implementations. However, we should not discard the possibility to increase their complexity by adding additional axonic or dendritic functionalities to them (such as signal gain needed to keep the amplitude of the pulse constant along the axon).

**Synaptors** play the role of synapses, controlling the connectivity between the neuristors. Non-volatile RS is an ideal effect to implement synaptic functionalities: the conductivity of the synaptor plays the role of the synaptic weight while its non-volatility mimics synaptic memory and plasticity. A thin (<100 nm) TMO layer sandwiched between two metallic electrodes is the easiest way to realize a synaptor although with recent lithography techniques planar synaptors have also been implemented.[38]

**Architecture** is a challenging issue that refers to the way in which neuristors and synaptors are wired. Current fabrication techniques limit hardware fabrication to 2D, making it complicated to achieve the huge 3D connectivity of biological systems. A partial solution to this is the implementation of cross-bar array geometries, such as shown in figure 3. In this geometry, neuristors are grouped in "layers", a row and a column, for input and output into a synaptor matrix. This architecture of layers is quite similar to that of software implemented neural networks and, to some extent, to real biological systems. Each neuristor has two terminals corresponding to the dendrite and the axon. The geometry is arranged so every "axon" of the pre-neuron layer intersects every "dendrite" of the post-neuron layer. Each intersection consists of a synaptor: A Metal/TMO/Metal non-volatile RS element playing the role of a synaptic connection. Multilayer, "deep", networks could be fabricated by connecting in series several of these arrays.

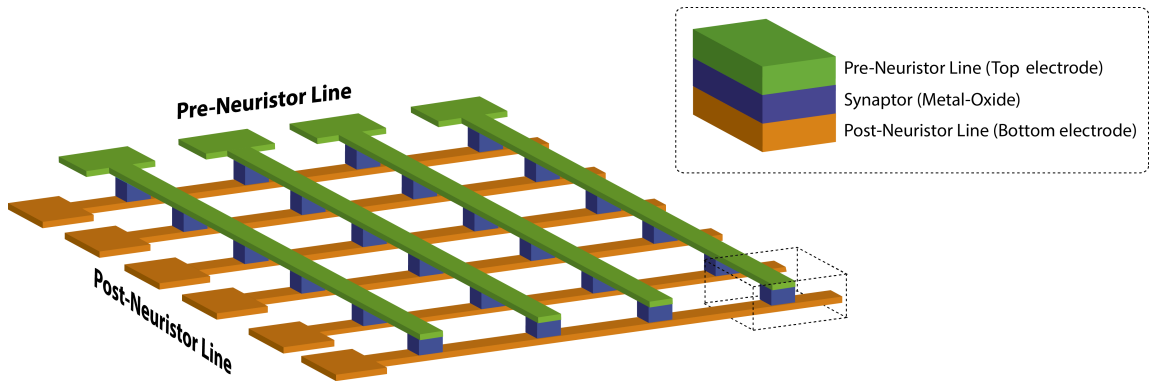

Figure 3: Cross-bar array architecture to implement hardware-based neural networks. It consists of a synaptor-matrix device in between a pre-neuristor layer and a post-neuristor layer.

The various needed neuromorphic functionalities may be implemented by a combination of resistive switching devices and metallic connections.[39] Resistive switching allows for implementation of basic neuronal and synaptic functionalities whereas the connectivity is provided by the metallic wires and pads. More detailed descriptions of successful neuromorphic devices implemented with resistive switching will be given in section VI.

**III. PHENOMENA**

III.1 Basics of Resistive Switching (RS)

Neuromorphic functionalities can be realized in practice by using and controlling the Resistive Switching phenomenon. The RS effect consists of a sudden change of the resistance of a system under the application of electric stress, such as applications of time dependent voltage, current and/or temperature. The microscopic origin and mechanism of the resistance change may be very different and, accordingly, qualitatively different for different type materials. As a first classification, there are two types of RS switching: *volatile* and *non-volatile*. In the former case the phenomenon is a sudden collapse of the resistance under the action of an electric stimulus, which *spontaneously* returns to the initial state.[40] This relaxation takes place gradually, a certain time after the electric action on the system is terminated. On the other hand, the non-volatile RS is a sudden change in resistance that is semi-permanent. It does not return to the initial state after the electric pulse is terminated, however, the change may be *reversed* by application of a new pulse. These non-volatile resistive switching behaviors may be further classified in two sub-categories: unipolar (also called non-polar) and bi-polar.[31] The former are systems where the resistive change can be reversed by the application of a second electric pulse of either the same or opposite polarity. In contrast, the bi-polar systems require that the second pulse be of the opposite polarity to the first one. It has become customary to call the pulse that produce the resistance change from high to low value as the "*set*" process, and the one that returns it to high resistance, the "*reset*" one. The current state of the system may be read with a "*read*" pulse, which is often a small bias current or voltage that should not affect the given resistive state.

Volatile resistive switching is generally observed in systems where a metal-insulator transition is observed (such as Mott insulators).[41] The resistance collapse occurs when part of the material changes from the insulating to the metallic phase as a result of an applied voltage. On the other hand, non-volatile switching arises from the redistribution of intrinsic defects or ions, mainly oxygen vacancies, which become mobile in the presence of strong electric fields. Their local density distribution changes and thus modifies the total (i.e. two-terminal) resistance of the material. In this section, we will describe the main microscopic mechanisms governing the different types of switching, which are shown schematically in figure 4.

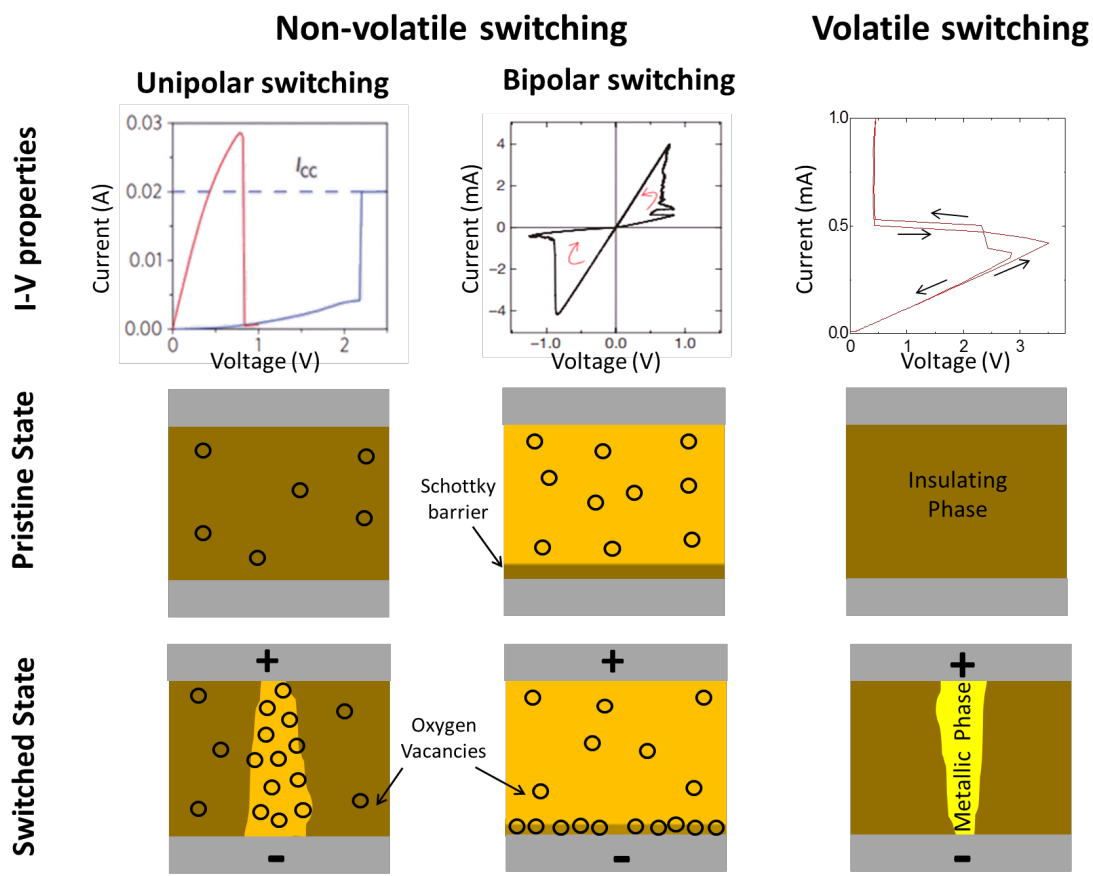

Figure 4: Schematic representation of the three main types of resistive switching: unipolar, bipolar and volatile (threshold) switching. Top panels show the current vs voltage characteristics of each resistive switching type. The grey regions denote metallic electrodes and the central region the RS material.

III.2 Origin of non-volatile resistive switching (NVRS)

The two qualitatively different types (unipolar and bipolar) of non-volatile RS systems, arise from different types of microscopic structural changes. The former emerges from filamentary structures that results from a soft dielectric breakdown or initial electro-forming process.[31] In contrast, the latter is observed in materials that are conductors and form highly resistive (Schottky) interfaces with the electrodes.

The unipolar systems typically are binary oxides (with some exceptions) that are initially highly insulating. On the other hand, the bipolar systems often are complex oxides and bad conductors, and do not require the soft dielectric breakdown.[42,43] Among these systems the cuprates and particularly the manganites have received significant attention.[43,44] This distinction is nevertheless not strict and highly oxygen deficient binary oxides are also often bipolar. For instance, $TiO_{2-x}$ has been shown to belong to the two categories, unipolar and bipolar, depending on fabrication details and the oxygen content.[45–47]

III.2.1 Unipolar RS

Unipolar (sometimes called non-polar) RS is typically observed in simple binary TMOs such as $TiO_2$, $NiO$, $CuO$, $HfO_2$, $Ta_2O_5$, etc.[26,32,33,48,49] These materials in their stoichiometric form are very good insulators. Typical devices are capacitor-like, where the dielectric is a thin film of the TMO. RS in these insulators is produced by *electroforming*, a process which produces a soft electric breakdown on the dielectric. Typically, MV/cm electric fields are needed, easily achievable with a few volts across ~100nm thin films. To prevent sample damage, one may use a load resistor in series, or impose a current compliance.

Electroforming produces a massive migration of ions (such as oxygen vacancies[50]), that leave behind filamentary structures of pure conductive transition metal. These filaments extend from one electrode to the other, have thickness of the order of nanometers[38] and conductance quantization has been reported.[51] The filaments may also result from a stable metallic phase of different stoichiometry. One example is filaments of metallic $Ti_4O_7$ in a matrix of insulating $TiO_2$, both members of the Magnelli series of titanium oxides.[52]

Once thin filaments are formed, the RS follows from the successive interruption and reconnection by electric field pulses. These two processes are, however, physically very different. Joule heating induces interruption of the filament, i.e. resistive change from the low- to high-resistance state (reset). When a current flow through the filament, the highest heating point is located at its thinnest part. Typical currents in the mA range, dramatically increase the local temperature (~900 K) rendering oxygen very mobile which produces local re-oxidation.[38] The conductive path is broken just at the local spot where it had a constriction. The reconnection process is also easy to understand. Joule heating is not important in this case, because the current is orders of magnitude lower in the high-resistance state. However, the local electric field across the two tips of the disconnected filament is very large. This strong electric field produces a new dielectric breakdown of the oxidized spot that had broken the filament. Oxygen vacancies migrate to the spot, the oxide is reduced and the filament reconnects. This produces the change from the high- to low-resistance state (set). Thus, the *reset* is driven by current (Joule heating), while the *set* is driven by voltage (electric breakdown). The typical time scales of these two phenomena are different, with the former being slower. Nevertheless, depending on the geometry of the system, nanosecond switching may be achieved. Clearly, the physical process involved in the switching are independent of the applied polarity, hence its unipolar character.

## III.2.2 Bipolar RS

Bipolar resistive switching is typically observed in complex oxides, such as $Pa_{0.7}Ca_{0.3}MnO_3$ (PCMO), $YBa_2Cu_3O_{7-x}$ (YBCO) , $SrTiO_3$ (STO), Zn-doped amorphous $SiO_x$ (SZO), etc., [26,31,41,53–55] but also in off-stoichiometry binary oxides, such as $TiO_{2-x}$, $HfO_{2-x}$, $Ta_2O_{5-x}$, etc. [33,46,48] These latter systems may also be realized by depositing a capping layer of the transition metal on top of its oxide. A qualitative difference with unipolar RS is that some bipolar systems may not need any electroforming[56], although still require a few initial electric cycles until they display reproducible switching. Unfortunately this advantage is not found in pristine highly insulating materials, which require a soft electric breakdown before becoming operational. This is a clear practical inconvenient and overcoming it remains an open challenge. The bipolar binary oxides are typically small devices, made of a few nm thick thin-film. In contrast, complex oxides can be synthesized in thin-film form using modern synthesis techniques or may be studied in bulk form. As we will discuss below in these latter systems the RS actually occurs at their Schottky interfaces with the metallic electrodes.[17,42,44,57]

The physical phenomena behind the RS are perhaps less clear than in the unipolar case. The systems are more complex and a theoretical description is a bigger challenge. Nevertheless, the unquestionable main common feature of all systems is that the RS involves the drift of ionic species under an electric field.[26] Unlike the previous case, bipolarity requires reversing the resistive state which implies back and forth migration of ions within the material. However, two relevant questions still stand: 1) How can ionic migration change significantly the resistance? 2) How can the switching speed be so fast?[53,26] The answers to these two questions arise from general constraints imposed on the physical mechanism. The relevant ionic drift is caused by the electric field across the dielectric material produced by the external voltage applied to the electrodes. The field is proportional to the local voltage drops across the system. Ohm's law, implies that the fields is proportional to the local resistivity $\rho(x)$, where x denotes a spatial coordinate across the dielectric between the electrodes. Thus, the highest fields appear at the highest resistivity regions. Since the bi-polar dielectrics are poor conductors they are expected to form Schottky interfaces with ordinary metal electrodes.[17,42,57] This Schottky interface is characterized by an electrostatic barrier that arises due to band bending to equilibrate the chemical potentials and the electrostatic forces.[58] These interfaces become depleted of carriers and thus the local resistivity of the dielectric is dramatically increased, although without a structural change. Thus, the RS effect in these two-terminal capacitor-like devices is dominated the highly resistive (Schottky) interfaces in an otherwise conductive material.[31] This observation is the key to the answers of the two questions posed above.

As we just described, upon the application of an external voltage the highest electric fields develop at the high resistance interfaces. As a consequence, the most significant ionic migration process takes place there. Moreover, the ionic drift induces local structural changes, leading to significant modulation in the (two-terminal) resistance of the devices, and hence, the RS[59] (figure 5). This helps understanding the fast switching speed of RS as the width of the Schottky barriers is typically a few nanometers and the ionic drift speed may be as high as ~m/s.[60] In fact, those high speeds may be easier to reach under strong electric fields along grain boundaries and by thermal effects.[53] Thus, at drift velocity of m/s it takes only nanoseconds to migrate across a Schottky barrier, which accounts for the fast-observed RS commutation speeds.

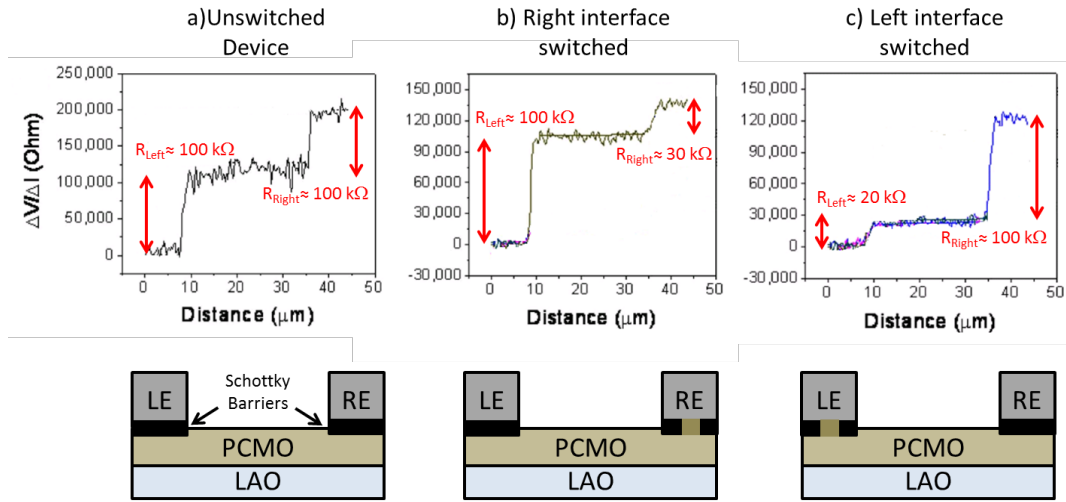

Figure 5. Top panels: Kelvin probe measurements (Adapted from reference [59]) of the local resistance for a) an un-switched device, b) when the right electrode has been switched and c) when the left electrode has been switched. The brown areas within the black interfaces indicate a region where the Schottky barrier has been lowered due to resistive switching. Bottom panels: Schematic representation of bipolar resistive switching.

To explain the physical change that takes place at the interface two possible phenomena have been invoked. Electrostatic modulation of the barrier height by in and out flow of ions in the interfacial traps.[31,61] Another possibility, consistent with the robust universality of the phenomenon, is the migration of oxygen vacancies which produce the positively charged defects in all TMOs. For instance, inducing a large number of vacancies may introduce carriers and turn an insulating binary oxide into a bipolar RS system.[46,47] Or, in more complex and conductive oxides, oxygen vacancies affect the electric transport, since they perturb the O-M-O bonds forming the conduction band of TMOs.[42]

The electrode choice has proven to be of critical importance. Metal/Insulator interfaces generally form Schottky barriers, while heavily doped insulators or semiconductors tend to form Ohmic contacts [62]. Inert electrodes, such as Pt, Pd or Au, do not change the stoichiometry of the oxide, creating a Schottky barrier interfaces that can be tuned with resistive switching. On the other hand, reactive metals such as Ti or Ta induce oxygen vacancies in the underlying oxide, doping the film and collapsing the barrier. This renders Ohmic interfaces that do not easily change their resistance when a voltage is applied [63]. We should point out that RS due to motion of the TM ions has also been reported in very thin films of the binary systems TaO$_x$, HfO$_x$ and $TiO_x$ (see also section IV.2.1).[64,65]

### III.3 Origin of volatile resistive switching (VRS)

The volatile resistive switching is a different physical phenomenon from the non-volatile one. It is observed in materials that feature a metal-insulator transition (MIT) as the temperature is lowered.[41] They evolve from a high temperature metallic phase to a low temperature insulating phase, with several orders of magnitude resistivity change. figure 6a shows the resistance as a function of the temperature for a $VO_2$ thin film, with the MIT clearly visible around 340 K.

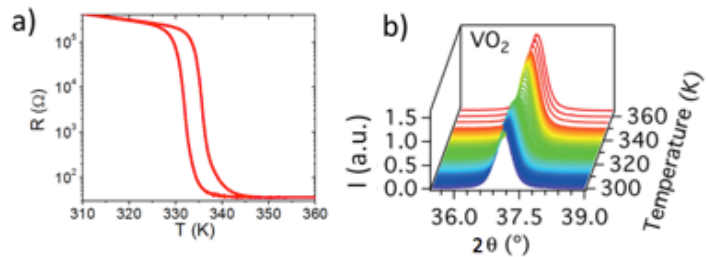

Figure 6: a) Metal insulator transition of a $VO_2$ thin film. b) Structural transition revealed by XRD. Adapted from [66].

Different mechanisms have been advanced to explain the origin of this transition. The most widely accepted is based on the mechanism proposed by N. Mott in the 40's. He argued that electron-electron repulsion due to Coulomb interaction would favor charge localization of the electrons in a partially filled band, rendering insulating an a priori metallic material. This insulating phase is commonly known as "Mott insulator". Repulsion energy between electrons becomes large enough to make double occupation energetically costly, limiting their hopping between neighboring sites. As temperature is increased, the stability of the Mott insulator electronic state may, however, become unfavorable as compared to the expected metallic state. The ensuing thermally driven metal-insulator transition is called a Mott transition. Hence, key to this phenomenon is the energy difference between two electronic competing states, which can be a very small energy scale.[67,68] Some materials are considered pure Mott insulators such as $GaTa_4Se_8$,[69] and Cr-doped $V_2O_3$,[70,71] since their $1^{st}$ order metal-insulator transition occurs in the absence of any symmetry breaking (either structural or magnetic). However, quite often the MIT is accompanied by a concomitant structural phase transition, and occasionally by low temperature magnetic ordering. Examples of these materials are $VO_2$ (structural transition shown in figure 6b), $V_2O_3$, $NbO_2$ or the nickelate family ($RENiO_3$).[41] Since there is a symmetry change the precise MIT mechanism maybe more elusive. This symmetry change implies two possible additional mechanisms: the Peierls and the Slater transitions. In a Peierls transition, the lattice structure change opens a gap at the Fermi level, creating insulating phase. In a Slater transition, the magnetic ordering changes the interaction potential seen by the electrons, modifying the effective symmetry of the lattice and opening the gap at the Fermi level. In a number of oxides ($NbO_2$, $VO_2$ and $V_2O_3$ [41]) the dominant mechanism for the MIT is still controversial and beyond the scope of this tutorial.

Independently of the physical mechanism behind the MIT, there is a surprising universality in the electrical transport observed in these correlated oxides. The most notable one is the possibility to induce the metallic phase by applying a voltage or a current in the material. Figure 7 shows the I-V characteristics of four different Mott insulators. There is a remarkable similarity between all of them: the resistance suddenly drops when a threshold voltage is applied, returning to its original value once the voltage is removed, yielding a reversible, volatile switching. This resistive switching corresponds to part of the sample undergoing a MIT into the metallic state, which produces a sharp resistance drop. The reason for this electrically triggered transition to take place is still controversial.[72] The simplest explanation is that current generated Joule heating locally raises the temperature above the MIT transition, inducing the resistive switching. There is some experimental evidence that this might play a role in the I-V characteristic of VO$_2$ and V$_2$O$_3$ small devices[73-75] depending on the gap size ratio[76]. However, such and explanation does not hold for the case of Cr-doped V$_2$O$_3$ that does not have a temperature-driven MIT. Thus, another explanation is that strong electric fields might destabilize the insulating phase, making the metallic state energetically more favorable. This may be revealed in different systems depending on the gap size of the device.[76,77] Both mechanisms are very difficult to differentiate from the I-V properties of the material, as both would create a threshold-like, volatile resistive switching. [72,76]

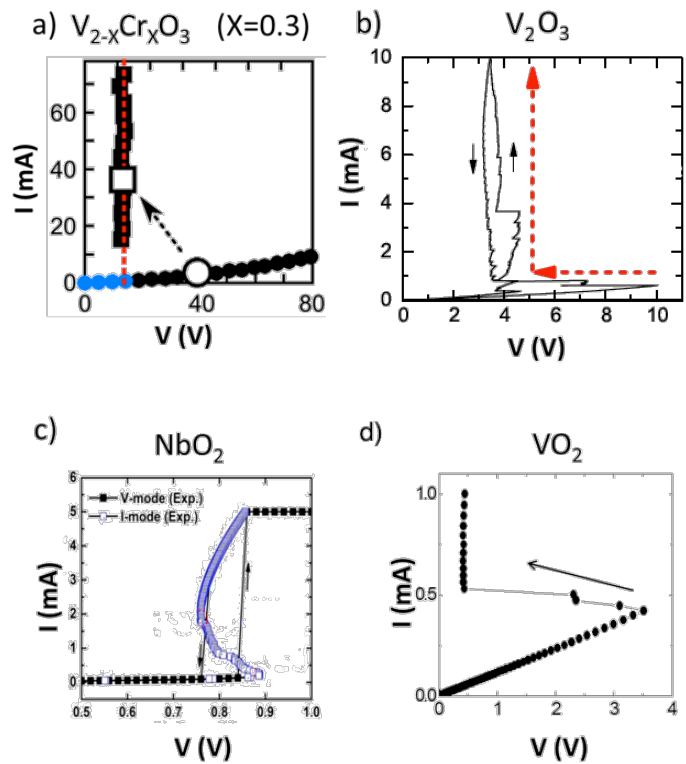

Fig 7. I-V characteristics of three different Mott insulators: a) V$_{2-x}$Cr$_x$O$_3$ (X=0.3)[78], b) V$_2$O$_3$ [74] c) NbO$_2$ [79] and d) VO$_2$[80]

Another mechanism of electrically triggered MIT is by filamentary percolation produced by electro-thermal domains (ETDs)[74], similarly to non-volatile resistive switching. This is expected independently of the triggering mechanism: Joule heating or electric field. Self-heating in such a device can be strong enough to cause an electro-thermal breakdown, i.e., a current and temperature redistribution. In any of those cases, the first part of the sample that becomes metallic will drag more current through it and locally create a more intense electric field. This will, for both mechanisms, promote longitudinal growth and eventually lead to the formation of filaments, as seen in figure 8 for $VO_2$ and $V_2O_3$.

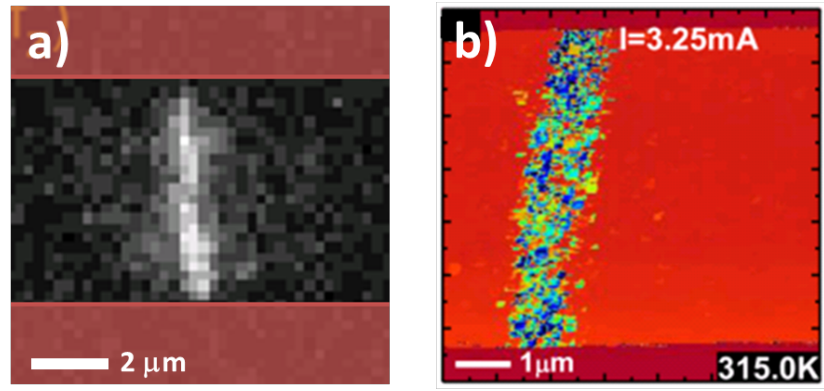

Figure 8. Experimental observation of a metallic filament in the current induced MIT in a) $V_2O_3$, obtained with low-T SEM image width corresponds to 10µm, from[74] and b) $VO_2$, obtained with near-field scanning microwave microscopy, from [81].

The characteristic times of the volatile switching, potentially could be several orders faster than the non-volatile switching. Indeed, pump-probe experiments have shown that the transition from insulator to metal can take place in the ps range. However, the implementation of these materials in real devices will impose a constraint in how fast the transition can be induced. The parasitic capacitance and the large resistance of these materials in their insulating state leads to RC constants of the order of the ns. [82] Moreover, depending on the device size, it will take a certain amount of time for heat to build up across the electrodes. This effect can be observed in figure 9 that shows results of a voltage pulse applied across a 100 nm Metal/$VO_2$/Metal structure. The transition does not take place immediately, but it takes around 10 ns for the metallic phase to be induced.

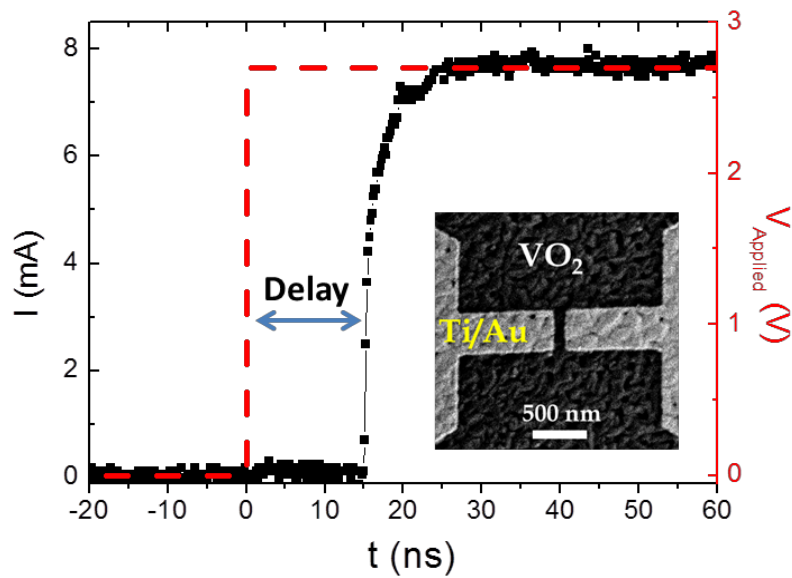

Figure 9. Current (left axis, black squared data) as a function of time when a voltage pulse is applied (right axis, dashed red line) to a VO$_2$ nanodevice (Inset). [80]

## IV. MATERIALS

IV.1 Why Transition Metal Oxides?

Resistive switching has been observed in a broad range of systems, ranging from phase change materials to solid solutions. We focus this review on one of the most promising materials systems, the transition metal oxides. This group of oxides has shown the highest versatility of all systems where this phenomenon is observed, as they may combine almost universal resistive switching with other unusual transport properties arising from electron-electron correlations.

TMOs display a host of remarkable phenomena and associated functionalities that include: superconductivity, colossal magnetoresistance, thermoelectricity, ferromagnetism, ferroelectricity, multi-ferroicity, metal-insulator transitions, etc. One may wonder why this bewildering variety of phenomena? What is special about these compounds? A few outstanding basic features arise. Chemically, the transition metals often present multiple valence states. These allow different coordination with oxygen which may lead to a variety of lattice structures. In addition, it also allows them to change the electronic occupation in the active shell (or band). These most relevant aspects lead to the unusual electric transport properties key to the understanding of the RS phenomena.

One of the most striking aspects of RS is its ubiquity. In figure 10 we show a periodic table where the transition metals whose oxides display non-volatile resistive switching are indicated.

Figure 10. Periodic table showing the elements whose corresponding binary oxide features non-volatile resistive switching. [33]

This ubiquity points to a rather general common origin for the RS phenomena and not some specific material-dependent effect. The common physical origin of non-volatile resistive switching relies on the fact that ions can drift in these oxides and, due to the multi-valence states of the transition metal, the resulting structures are metastable. One important feature of this phenomenon is that it relies on creating defects and inducing their migration. Thus, another common feature of all these systems is that the RS commutation between a high and a low resistive state displays a certain degree of variability. Therefore, non-volatile RS is not induced by a *bona fide* thermodynamic phase transition but rather due to different metastable structural states.

While not belonging to the family of TMOs, it is also worth to mention the important family of silicon oxides, which have displayed similar NVRS phenomena as described here and. From a practical point of view, these materials have the advantage of excellent CMOS compatibility.[83–86]

Regarding the volatile RS effect, transition metals also play a prominent role and we may consider the phenomenon universal, but in a different sense. Volatile switching is generally observed in Mott systems, which exhibit a metal to insulator transition. Conventional band theory predicts that a half-filled band should be a metal. However, the Mott insulator state may arise due to the localization of electrons due to strong on-site Coulomb repulsion. Qualitatively, in a monovalent system with narrow orbitals, the energetic cost of doubly occupying a site may be so high that the system chooses to minimize its energy by keeping one electron localized at each atomic site. The possibility of either delocalizing or localizing an outer shell electron seems to be a quality of transition metals. The idea is that *d*-orbitals are neither too big, as *s* and *p*'s that yield bands, nor too small, as *f*'s that yield local magnetic moments. Instead *d*-electrons "hesitate" between these two possibilities and may adopt either one depending on the detailed surrounding environment.

A comprehensive table with various metal oxides in which different types of resistive switching in observed is presented below (table 1). This table is intended as illustration and guidance for the reader rather than an exhaustive list.

| Oxide | Electrode Top / Bottom | Polarity | $E_{Set}$ (kV/cm) | $E_{Reset}$ (kV/cm) | Reference |
|---|---|---|---|---|---|
| MgO | Pt/Pt | Bipolar | ~1000 | ~400 | [87] |
| | Pt/Pt | Unipolar | ~1000 | ~400 | [87] |
| Al$_2$O$_3$ | Ti/Ti | Bipolar | ~250 | ~250 | [88] |
| SiO$_x$ | Poly-Si/Poly-Si | Unipolar | ~1000 | ~1000 | [83] |
| TiO$_2$ | Al / Ru | Bipolar | ~300 | ~150 | [45] |
| | Al / Ru | Unipolar | ~300 | ~150 | [45] |
| | Ni / Cu | Bipolar | ~100 | ~80 | [89] |
| | Ni / Cu | Unipolar | ~100 | ~80 | [89] |
| | Pt / Pt | Unipolar | ~200 | ~400 | [52] |
| VO$_2$ | Ti / Ti | Irreversible | ~800 | - | [90] |
| | TiN / TiN | Volatile | ~50 (300 K) | - | [82] |
| V$_2$O$_3$ | Ti / Ti | Irreversible | ~1200 | - | [90] |
| | Ti / Ti | Volatile | ~300 (10 K) | - | [75] |
| CrO | Pt / TiN | Bipolar | ~460 | ~400 | [91] |
| MnO | Ti / Pt | Bipolar | ~90 | ~100 | [92] |
| Fe$_3$O$_4$ | Pt / Pt | Bipolar | ~400 | ~500 | [93] |
| CoO | Pt /Pt | Unipolar | ~130 | - | [94] |
| NiO | Pt / Pt | Unipolar | ~120 | - | [94] |
| NiO | Pt / Pt | Bipolar | ~200 | ~100 | [95] |
| Cu$_2$O | Pt / TiN | Bipolar | ~80 | ~80 | [96] |
| | Pt / Pt | Bipolar | ~50 | ~30 | [96] |
| | Pt / STO | Bipolar | ~100 | ~100 | [96] |
| ZnO | TiN / Pt | Bipolar | ~300 | ~400 | [97] |
| GaO$_x$ | Pt/Pt | Bipolar | ~200 | ~100 | [98] |
| | Pt/Pt | Unipolar | ~200 | ~100 | [98] |
| GeO$_x$ | Cu/W | Bipolar | ~400 | ~200 | [99] |
| ZrO$_x$ | Pt / p$^+$-Si | Unipolar | ~400 | ~300 | [100] |
| NbO$_2$ | TiN / TiN | Voltatile | ~1000 (300 K) | | [101] |
| Nb$_2$O$_5$ | Pt / p$^+$-Si | Unipolar | ~1000 | ~500 | [102] |
| MoO$_x$ | Pt-Ir / Pt | Bipolar | ~150 | ~100 | [103] |
| | Pt-Ir / Pt | Unipolar | ~150 | ~50 | [103] |
| HfO$_2$ | Ti / Pt | Bipolar | ~250 | ~250 | [104] |
| | Ti / Pt | Unipolar | ~500 | ~250 | [104] |
| | Pt / TiN | Bipolar | ~3000 | ~400 | [105] |
| | Pt / TiN | Unipolar | ~3000 | ~200 | [105] |
| TaO$_x$ | Ti / Pt | Bipolar | ~400 | ~120 | [104] |
| | Ti / Pt | Unipolar | ~400 | ~150 | [104] |

| | Ta / Pt | Bipolar | ~1000 | ~200 | [106] |
|---|---|---|---|---|---|
| | Ta / Pt | Bipolar | ~300 | ~300 | [107] |
| **ZnO$_x$** | ZnS/ZnO$_x$ | Bipolar | ~22 | ~22 | [108] |
| **WO$_x$** | TiN / W | Unipolar | ~300 | ~600 | [109] |
| **CeO$_x$** | Al/Pt | Bipolar | ~100 | ~500 | [110] |
| **Gd$_2$O$_3$** | Pt/Pt | Unipolar | ~200 | ~100 | [111] |
| **Yb$_2$O$_3$** | Pt/TiN | Bipolar | ~500 | ~500 | [112] |
| **Lu$_2$O$_3$** | Pt/Pt | Unipolar | ~300 | ~150 | [113] |
| **STO** | Au / Au | Bipolar | ~0.1 | ~0.1 | [53] |
| **PCMO** | - | Volatile | ~10 (20 K) | - | [61] |
| **PLCMO** | Ag / Ag | Bipolar | - | - | [56] |
| **BFO** | Cu / Pt | Bipolar | ~150 | ~150 | [114] |
| **YBCO** | Au / Au | Bipolar | - | - | [115] |
| **SZO** | Au / SRO | Bipolar | ~15 | ~15 | [54] |

*Table 1. Examples of oxides in which resistive switching is observed. The type of switching and set and reset fields are indicated.*

## IV.2 Other systems displaying resistive switching

Although this review is mainly focused in TMO, other material systems show resistive switching. Here we will give a brief introduction to two of such systems: conductive bridge memories and phase-change materials.

### IV.2.1 Cation Drift systems

Conductive Bridge memory (CBRAM) also known as electrochemical metallization memory (ECM), or "cation" systems are produced by the drift of positive metallic anions (as opposed to oxygen vacancy cations in the case of TMO) [23,32,116–118]. These devices are generally composed by two different electrodes: an electrochemically active (generally Cu or Ag) and an inert (such as Pt, Au, Cr or W) electrode. These two electrodes are separated by an insulator such as a solid electrolyte with good ionic conductivity (Ag or Cu doped sulfides or chalcogenides: Ag$_2$S [119,120], GeS$_x$ [121], GeSe$_x$ [121,122] or pure dielectrics (a-Si [123,124], ZrO$_2$ [125], HfO$_x$ [126], SiO$_2$ [123,127] or Al$_2$O$_3$ [128])

A positive voltage applied to the active electrode oxidizes the metal (Ag or Cu), producing positive metallic ions that drift through the insulator pushed by the electric field (Figure 11, point B). When the anions reach the inert electrode, they are reduced, forming a metallic deposit which grows from the inert electrode towards the active electrode, as shown in point C of Figure 11. When it reaches the opposite electrode, a sudden resistance drop takes place (Point D). Reversing the polarity reverses the effect: Ag or Cu will be oxidized at the inert electrode interface and pushed in the opposite direction, breaking the conductive path between electrodes (Point E). This process creates a bipolar NVRS. [23,32,116–118]

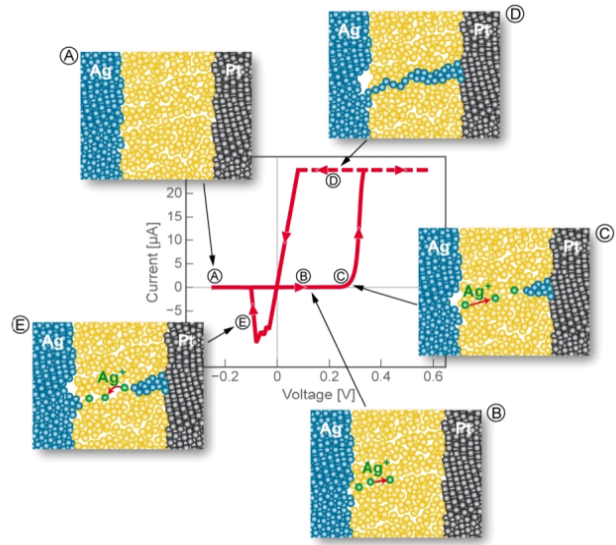

Figure 11. Schematic representation of filament formation and bipolar resistive switching in CBRAM. Adapted from reference [117].

The RS properties of these systems are strongly dependent on the morphological properties of the insulating film.[23,117,123] Good ionic conductors and granular films facilitate ion motion, leading to long filamentary/dendritic structures and low forming voltages.[123,125] On the contrary, densely packed films (such as Si) limit cation supply and induce the formation of metallic nanoclusters inside the insulating matrix.[125,126] These clusters deform or split when an external voltage is applied, modifying the resistance of the device.[123,126]

Volatile resistive switching has been recently observed in this type of systems. Wang *et al*.[126] fabricated cation devices consisting on Ag nanoparticles embedded in different oxide matrixes and sandwiched between two inert electrodes. These devices spontaneously recover their high resistance once the applied voltage is removed, typically in a ms timescale (Figure 12a). Using TEM, it was argued that interfacial tension might cause this effect: the nanoparticles merge into larger clusters once the electric field is removed, breaking the conductive path (Figure 12b). This shows that CBRAM dynamics is a complex phenomenon and several mechanisms (electrochemical, thermal and mechanical) might be at play.[126]

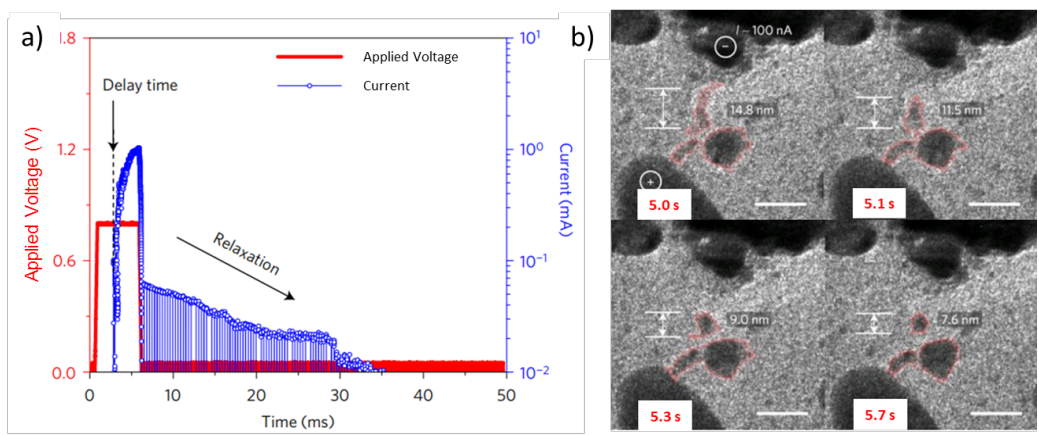

Figure 12. Volatility in a CBRAM device. a) Current as a function of time: the device returns to its insulating resistance value after a few ms. b) HRTEM observation of the relaxation process: smaller Ag nanoparticles coalesce into larger ones when the external voltage is removed at t=5.0 s. Adapted from reference [126].

IV.2.2 Phase-Change Materials (PCM)

The basic feature that enables RS in this type of materials is that they can exist in two very different states: low resistance crystalline phase, or high resistance amorphous phase. Switching between both is possible by heating the sample and carefully controlling the cooling rate.[129–133] This can be done either optically[131,134–137] or electrically,[129–133,138–142] making phase change materials versatile for applications in photonics as well as in electronics. The amorphous phase can be induced by applying a short, high voltage pulse to the sample: intense enough to drive it over its melting point and sharp enough to allow thermal quenching into the amorphous state (RESET pulse in Figure 13). The crystalline phase can be recovered by applying a longer, lower voltage pulse that allows recrystallization by annealing (SET pulse in Figure 13).

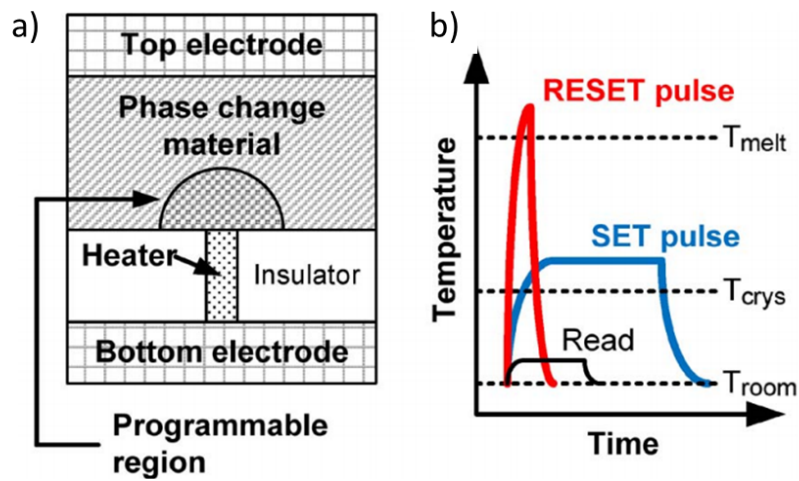

Figure 13. a) Typical structure of PCM devices. b) Schematic representation of the SET, RESET and read pulses used in PCM. Adapted from reference [132].

Materials used for this type of memory should in general meet two requirements: large crystalline/amorphous resistivity ratios and fast re-crystallization times. [129–133] Semiconducting alloys in pseudo-binary line between GeTe and $Sb_2Te_3$ (such as $Ge_2Sb_2Te_5$) show crystallization times in the $10^{-8}$ s range and up to 5 orders of magnitude resistivity change.[133,135] Their discovery in 1987[135] allowed the development of commercial technologies based on PCM, such as rewritable optical storage (CD-RW). [131,133] Three other systems with fast crystallization times were later discovered: $Sb_2Te$ doped with Ag, In or Ge[136]; GeSb[137] and $GeTe_x$ alloys.[138,139] The typical RS currents, voltages and characteristic timescales are comparable to those observed in TMO [140], making PCM a direct competitor for neuromorphic computing implementations.[141,142]

## V. MODELS

So far, we have presented the different ways that resistive switching may appear and the materials where the phenomena are observed. In this section, we will briefly review some of the theoretical models that have been proposed to explain the experiments. Table 2 shows the different types of resistive switching, together with their origin and the most notable models that explain their phenomenology (Please note that conductive bridge memories and phase change materials are not included in the table).

| | **Non-Volatile Resistive Switching** | | | **Volatile Resistive Switching** | |
| | **Unipolar** | **Bipolar** | | | |
|---|---|---|---|---|---|
| **Origin of Switching Mechanism** | Oxygen vacancy generation by electroforming | Oxygen vacancy drift induced by an electric field | | Metal-insulator phase transition | |
| **Materials** | Binary oxides (mainly) | Complex oxides, off-stoichiometry binary oxides. | | Mott insulators | Mott insulators |
| **Set Mechanism** | Metal or metal-phase filament formation induced by electric field | Oxygen vacancy drift in the bulk | Defect drift in/out interface | E-field induced transition | Heating induced transition |
| **Reset Mechanism** | Joule heating induced reoxidation of filamentary weak link | Oxygen vacancy drift in the bulk | Defect drift in/out interface | Relaxation into insulating phase | Relaxation into insulating phase |
| **Models** | Random Circuit Breaker | Memristor | V Enhanced O Drift | R Network E-Field Induced | R Network Heating Induced |

*Table 2. Characteristics of the different types of resistive switching. V Enhanced stands for Voltage Enhanced, O Drift stands for Oxygen Drift and R Network stands for Resistor Network. Note that conductive bridge memories and phase-change materials are not included in the table.*

V.1 Non-volatile resistive switching (NVRS)

Below we describe briefly some of the theoretical models proposed to explain the non-volatile RS effect which provide numerical support to the qualitative descriptions discussed above. As done in the previous sections, we shall make the distinction between unipolar and bipolar systems.

V.1.1 Unipolar NVRS

The Random Circuit Breaker (RCB) model consists of a resistor network[143] (figure 14) in which each unit takes one of two resistance states (high) $r_h$ or (low) $r_l$, with $r_h \gg r_l$. The model assumes that each unit can undergo a transition, following the rule $r_h \to r_l$ if V >

$V_{on}$ and $r_l \rightarrow r_h$ if $V > V_{off}$. The value $V$ is the local voltage drop at each element and $V_{on}$ and $V_{off}$ are threshold voltage values, with the condition that $V_{on} \gg V_{off}$.

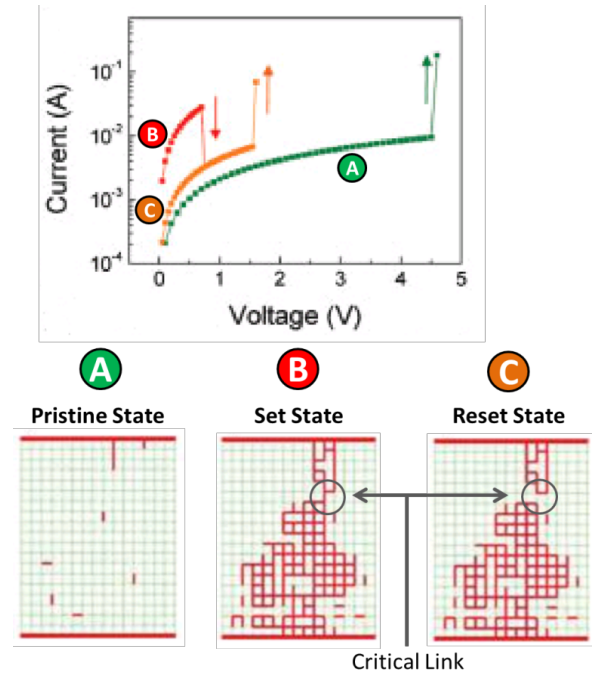

Figure 14. Random circuit breaker model, adapted from [143]. Top panel shows the simulated I-V characteristics of the model. Bottom panels show the distribution of metallic domains within the oxide matrix.

The applied voltage is initially ramped up, and eventually an avalanche of transitions takes place (figure 14). At this point the system has been "electroformed" as low resistance links have percolated. The unipolar RS effect is produced as the applied voltage is ramped up again from zero. The local voltage-drop increases until a low resistive element reaches the lower threshold $V > V_{off}$ ($\gg V_{on}$) and switches back to the $r_h$ state. Eventually, percolation is lost at a last "critical" link (figure 14 panels B to C). Then, the total resistance of the network jumps to a high resistance $R_{high}$. For a sufficiently large difference between $r_h$ and $r_l$ most of the applied voltage drops across the final link. This then switches to the low value $r_l$ when the external voltage is increased again. Thus, after the initial electroforming event, the successive hi-lo resistive switching is concentrated on the on/off switching of a critical link.

The RS effect can also be modeled by solving the differential equations that describe the electric and thermal transport of a conductive filament surrounded by an insulating matrix [48]. The model further assumes that the filament dissolves by out diffusion of its conductive elements into the insulator. The diffusion process is assumed to be activated which leads to a diffusion velocity with an Arrhenius dependence $V_D = V_{D0} \, exp(-E_a/k_B T)$, where $E_a$ is an activation energy. The electrodes are assumed to be the heat sinks. As the temperature increases in the center of the filament the filament dissolves, which further increases the temperature by the local Joule heating until the filament breaks.

### V.1.2 Bipolar NVRS

An early qualitative model of bipolar systems[144] anticipated several features, which provided useful guidance. The model assumed a relevant role played by the electrode-material interfaces and the ionic migration which appears in many of the later models.

The *memristor* model,[46,145] captured a great deal of attention (figure 15). It was introduced as the "missing element", arising from symmetry considerations of electric circuit theory. This simple model was formulated to rationalize bi-polar RS experiments in $TiO_{2-x}$ devices.

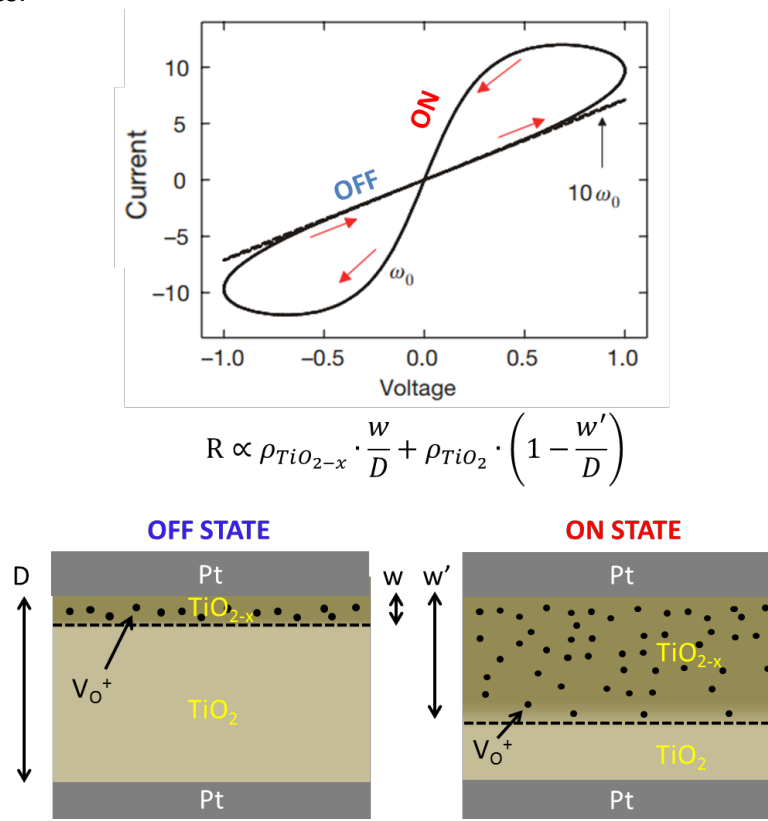

$$R \propto \rho_{TiO_{2-x}} \cdot \frac{w}{D} + \rho_{TiO_2} \cdot \left(1 - \frac{w'}{D}\right)$$

Figure 15. Top panel: Simulated I-V characteristic of the memristor model at two different ac frequencies. Bottom panels: Schematic representation of oxygen vacancies distribution for the ON and OFF states. w and w' denote the size of the oxygen depleted layer, while D is the distance between electrodes. Adapted from [46].

It assumes that the dielectric part of the device has two regions: a low conductivity, doped $TiO_{2-x}$ of length *w*, and an insulating pure $TiO_2$ of length D-*w*, where D is the distance between the electrodes. Thus, the model assumes two series resistors of high and low resistivities $\rho_{off}$ and $\rho_{on}$. This leads to a device resistance R = $R_{off}$ *w*/D + $R_{on}$ (1-*w*/D), where $R_{off}$ and $R_{on}$ are the resistances of the pristine and oxygen depleted films, respectively. The crucial model assumption is that the applied voltage across the device produces ionic (oxygen vacancies) drift, which leads to the shift of the location of the two-region interface. Thus, the device is no longer Ohmic and the I-V characteristics becomes non-linear. This type of memristive model applies to small nano-size systems, where the insulating part is small enough and the electric fields that can develop are large enough to promote ionic drift of defects. This simple model produces the generically observed pinched hysteresis loop in the *I-V* characteristics (figure 15). This type of memristive model can be extended to incorporate the temporal evolution of *w* with the applied voltage. This leads to a full dynamical model which, in principle, may describe the behavior of the memristor upon application of arbitrary pulse and ramp voltage shapes. The drawback of this model is the need to measure and fit an extensive set of data with ad-hoc mathematical expressions containing various free parameters.[146]

Another bipolar model is the so-called voltage enhanced oxygen vacancy drift (VEOD) model[42] (figure 16) formulated to describe an unusual RS effect in Pt/PCMO/Pt devices.[147] This behavior demonstrated the key-role played by the highly resistive Schottky interfaces and their complementary behavior in symmetric devices.

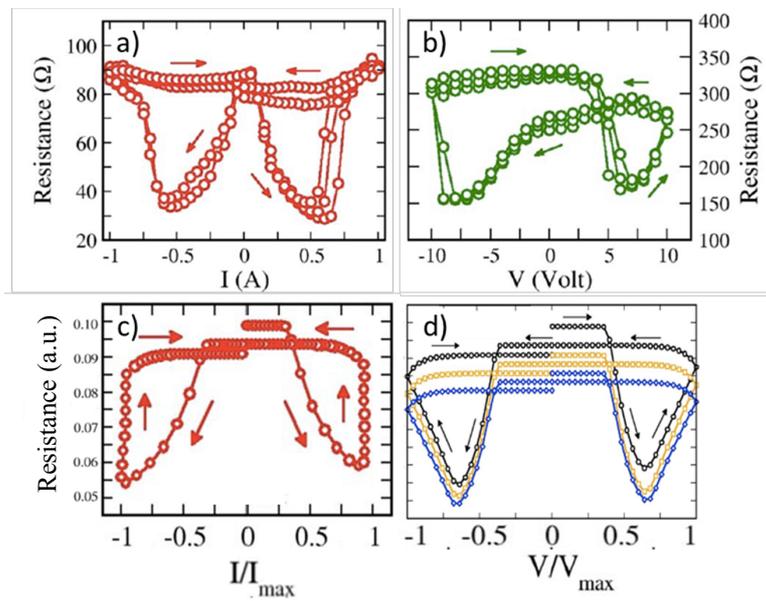

Figure 16. RS experimental data in manganite a) and cuprate b) devices driven by current and voltage, respectively. RS simulation using the VEOD model in a current c) and voltage (d) controlled simulation. Adapted from [42].

It should be mentioned, that models of voltage-enhanced drift[42,46] naturally lead to the possibility of shockwave propagation of the ionic motion during resistive transition.[148,149]

V.2 Volatile resistive switching (VRS)

The volatile resistive switching originating from a destabilization of a Mott insulator under a strong electric field possess a great theoretical challenge[77,150] with limited contact with experiments. An intriguing consequence from the theoretical studies is that the electric breakdown of a Mott insulator is predicted to occur at MV/cm electric fields, while experimentally is observed for fields of only kV/cm.[37,151] Clearly modeling these systems requires considering realistic situations which include inevitable extrinsic and intrinsic disorder. Further insight may be obtained from phenomenological models to describe the volatile RS effect. The main proposed models are either based on resistor networks or on systems of differential equations that couple the thermal balance with the electric transport.

## V.2.1 Resistor network

The resistor network model shown in figure 17 was introduced to describe the universal volatile RS observed in VO$_2$,[152] V$_2$O$_3$,[74,153] Cr-doped V$_2$O$_3$, NiSe$_{2-x}$S$_x$ and GaTa$_4$Se$_8$.[37] In this model, each resistor of the network can be in one of two resistance states: metal or insulator. The probability to be in the metallic or insulating states is given by a free energy that schematically represents the first order nature of the phase transition.[71] There are two basic variations of this model, depending on the triggering mechanism of the MIT: E-field[37] or Joule heating.[74] In the case of E-field driven transitions, an extra term is added to the free energy to take into account the field induced destabilization of the insulating phase.[71,77]

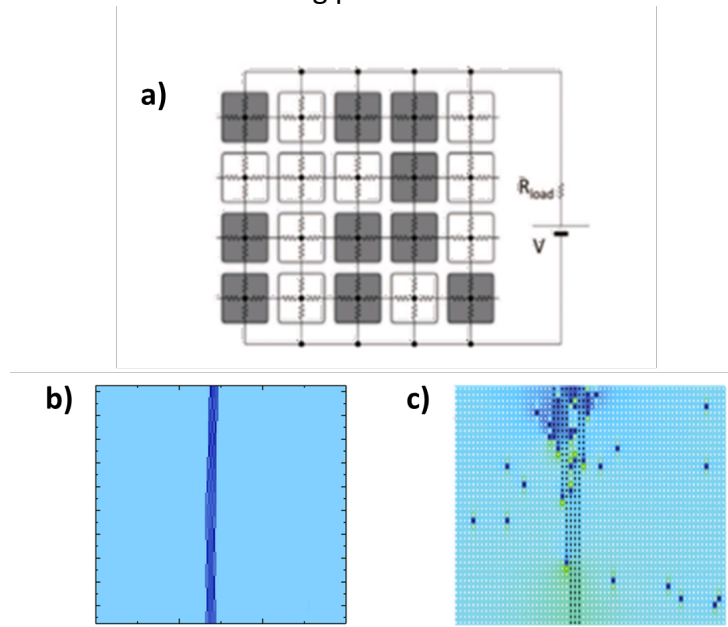

Figure 17. a) Typical resistor network configuration used for simulations of resistive switching in Mott materials. Each cell can be either metallic or insulating (black and white). Bottom panels show two simulations in which filament formation is observed for: b) a Joule heating induced transition (adapted from[74]), and c) an E-field driven transition (adapted from [37]).

The model assumes that the insulator is more stable at zero field than the metal and that they are separated by an energy barrier. The resistive cells of the network may change their state following an Arrhenius type thermal activation with the further crucial assumption that the energy of the insulator is affected by the local electric field acting on the resistive cell. This model captures qualitatively the experimental I-V characteristics and the rapid decrease with applied field of the delay time $\tau_d$ of the resistance collapse.

## VI. IMPLEMENTATIONS

Resistive switching in transition metal oxides, strongly correlated, and phase-change systems may enable novel neuromorphic electronic functionalities. As discussed above, the *non-volatile* RS enables "synaptic" functionalities, while the *volatile* RS enables "neuronal" functionalities. These two functionalities are the essential components of any neural network. This Chapter describe some of the recent progress in the implementation of such phenomena into real devices with computing capabilities.

## VI.1 Synaptic functionalities

The simplest implementation of a non-volatile synaptic function is to control the *synaptic weight* i.e. the electric coupling between two electronic neurons. This is achieved by using non-volatile RS to tune the resistance (or conductance) of the device. As in biological systems that modify the brain synaptic connections when a new memory is acquired, this is may be accomplished in a neuromorphic system by changing the resistance of the RS device. In practice, this can be achieved using electric pulses that selectively target different synapses in the neural circuit.

Besides the simple modulation of a synaptic intensity, other more elaborate synaptic functionalities are also being currently proposed and investigated. STDP functionality is at the root of learning algorithms for spiking neural networks. Figure 18 shows a concrete realization in a diffuse (SiO$_X$:Ag) electrode and a drift TiO$_2$ memristor-based synapse. [126]. STDP is obtained by careful design of the electric pulses emitted by the neuristors (see figure 18a). This way, the voltage across the memristor controls the synaptic weight change, by the relative timing between pulses.

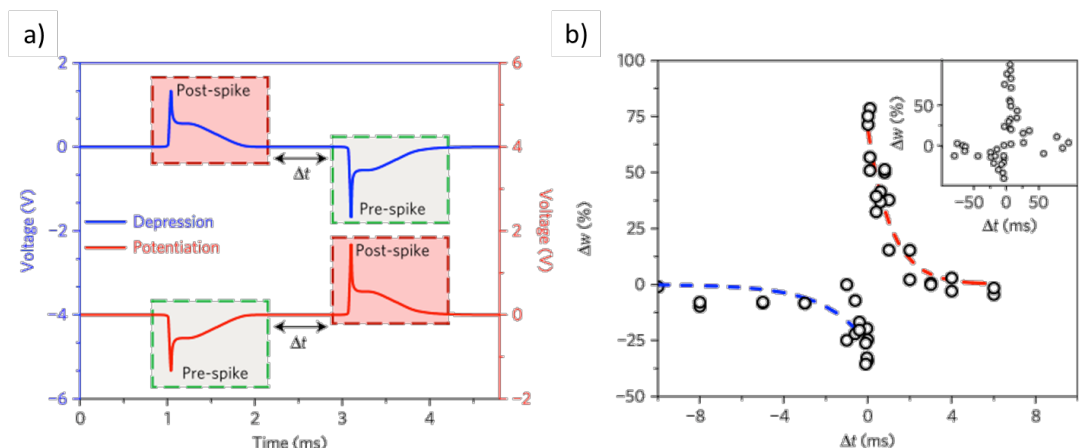

Figure 18. a) Schematic of the pulses applied to the combined device for STDP demonstration. The long low-voltage pulse in each spike turns the diffusive memristor ON, and the short high-voltage pulse switches the drift memristor. When the post-spike precedes the pre-spike, the device is reset (depressed), and when the pre-spike precedes the post-spike, the device is set (potentiated). The timing (⇆t) between the two spikes determines the voltage drop across the drift memristor. b) Demonstration of STDP. Synaptic weight change of a memristor as a function of the time difference between (pre and post) spikes of the system. The inset shows the qualitative agreement with a biological system. Figure adapted from [126].

## VI.2 Neuronal functionalities

As we described above, the neuron-like functionalities enabled by volatile RS can be realized using Mott insulator materials. Here we shall briefly describe some of these recent implementations.

### VI.2.1 Generation of electrical spikes

One of the goals of a neuristor implementation is to emulate the generation of an action potential. This could be mimicked[35] using the negative differential resistance of NbO$_2$ (figure 19a), as it undergoes an insulator to metal transition upon heating. This was accomplished by using two NbO$_2$ volatile memristors coupled to two capacitors (figure 19b) following the conceptual model of Hodgkin and Huxley (HH). This biological model describes the response of Na and K channels in the neuron membrane. The RS threshold behavior of the RS in the NbO$_2$ memristor, together with the RC constant of the capacitors mimics many key features of biological neurons such as spiking, signal gain and the refractory period (figure 19c).

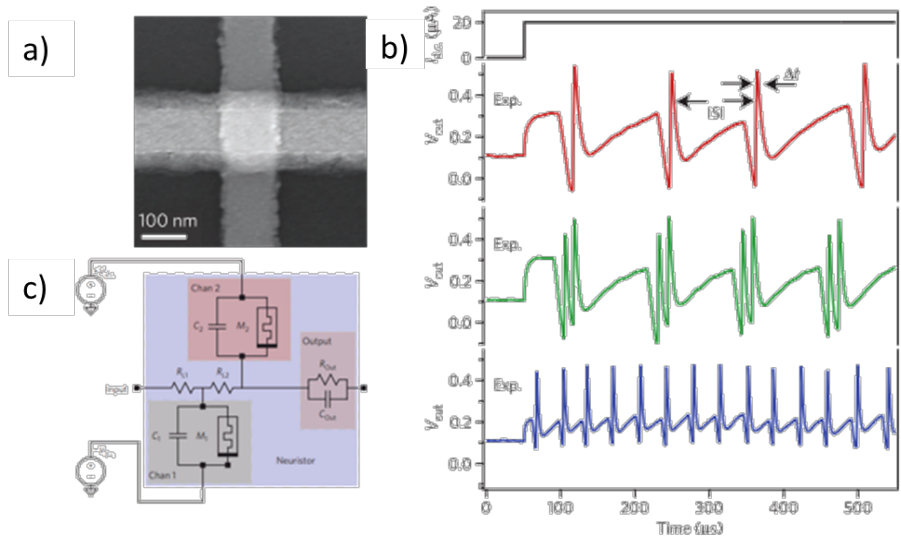

Fig. 19 Neuristor implementation based on NbO$_2$ memristors. a) SEM of a typical memristor b) circuit implementation of the neuristor to simulate a two-channel model. The device uses two nominally identical NbO$_2$ Mott memristors based c) input and output of the neuristor based-device. Top curve shows the dc current input. Bottom curves shows the device response for different $C_1/C_2$ (see b) configurations which emulate different frequency neuron responses.[35]

The basic feature of the circuit is an "all-or-nothing" behavior, namely, a small output signal until the input overcomes a given threshold value that produces the output voltage to develop a large spike (see figure 19c). In addition, this device may also produce regular trains of voltage spikes upon application of a constant dc current input (see figure 19c). Interestingly, the number, frequency and shape of the spikes can be tailored through tuning of the associated capacitors. This property may be appealing for implementation of systems with different types of neural coding schemes.

## VI.2.2 Leaky-integrate-and-fire

Another well-known property of neurons is the leaky-integrate-and-fire (LIF). In this model, the neuron acts as an as lossy *integrator* of the electric signal that arrives to its body (via its dendrites). At a certain integration threshold, the neuron *fires* a spike that travels down its axon to act on other "downstream" connected neurons. Quite remarkably, this functionality can also be implemented with Mott RS devices. In this case, the volatile RS behavior is used as described in section III.2. A 3D Mott insulator with small energy gap undergoes a resistive transition within a short delay time $\tau_d$ upon application of an above-threshold voltage $V > V_{th}$. The time delay depends on the strength of the applied V. When the input voltage is terminated, the resistance of the systems spontaneously recovers its original value within a characteristic time $\tau_r$. The key feature that leads to the LIF behavior is the response of the Mott RS upon application of a train of voltage pulses.[78,154] A single above-threshold pulse applied for $\tau_{on} < \tau_d$, does not produce the resistive collapse. A subsequent pulse applied after a time $\tau_{off}$, leads to two possibilities. If $\tau_{off} > \tau_r$ the system relaxes any change induced by the initial pulse, so the second pulse will also fail to induce a resistive collapse. However, if $\tau_{off} < \tau_r$ the effect induced by the first pulse will not be totally erased by the time the second pulse arrives. Further pulses produce a cumulative (leaky-integration) effect, eventually producing a resistive collapse and an ensuing current spike (fire).[37]

The evolution of the resistor network model (section V.2) upon an input of a train of pulses can be described by a set of differential equations analogous to the biological LIF model.[37] Moreover, given the strength and frequency of an incoming train of pulses (spikes), a mathematical formula has been derived which predicts the number of input spikes that will produce the fire event (i.e. resistive collapse).[37] This provides a remarkable functionality similar to biological neurons; the response of the system is faster when the incoming train of spikes is either more frequent and/or of stronger intensity (see figure 20)

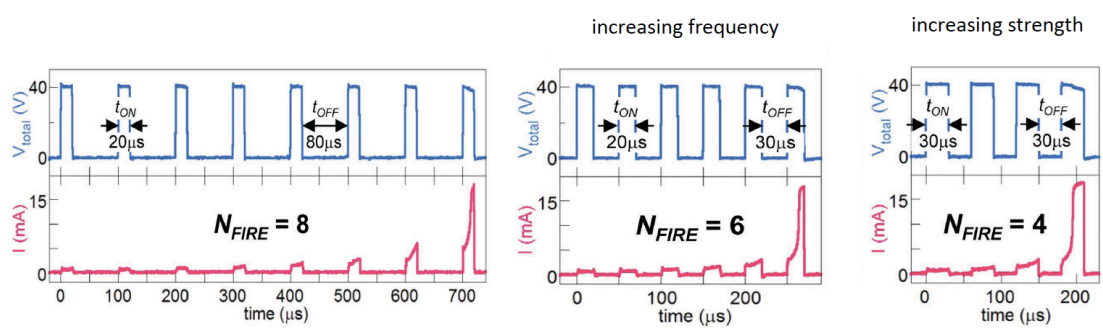

Figure 20. Experimentally observed decrease of the number of incoming spikes to set a fire event (resistance collapse) upon increasing pulse frequency or width, in agreement with the LIF model. Adapted from reference [35].

## VI.3 Crossbar architecture

High connectivity among neurons and selective access can be achieved partially by the regular crossbar architectures. The memristor crossbar arrays offer advantages[155] in machine learning implementations due to two key features[156]. First, the possibility to implement matrix operations based on the *synaptic weight* in a two-terminal device. Second, the option of online learning by using a sequence of pulses[155] Typical memristor crossbar arrays architectures are shown in figure 21. Each node is a two terminal memristive device. Inputs are connected to the network rows and outputs to the columns. The electric pulses at the input control the memristors (synaptic) weights. These designs are used in several implementations for data clustering techniques like principal component analysis in large-scale datasets[155] and pattern recognition.[157]

An important disadvantage of the crossbar implementations is the "sneak-path" problem due to an excess of current ("leakage") when one of the nodes is in the high-resistance state. When reading such a state, the current can flow through an unintended path that includes low resistance nodes giving an incorrect readout. Such possibility is depicted in figure 21. Several ways to overcome these problems have been proposed with marginal success. For example, grounding the floating terminals to divert the current to the ground,[158] a multistage reading protocol,[159] an unfolded architecture using a read lane for individual memristors (figure 21), diode or transistor gating[160] and using the non-linearity current-voltage characteristic of metal-oxide memristors[161,162] have not fully solved the problem.

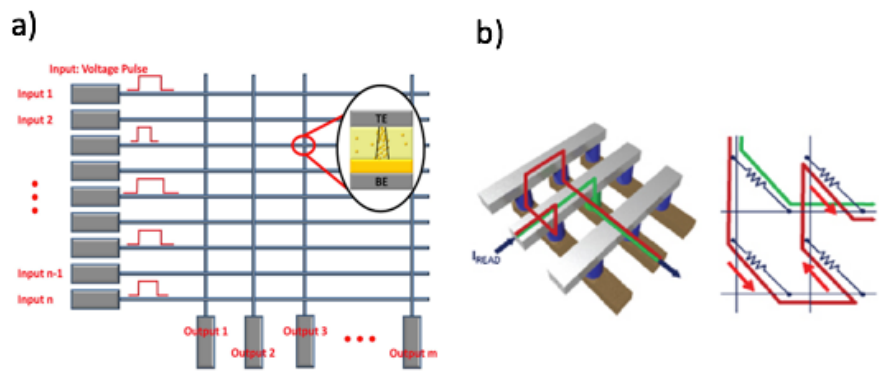

Figure 21 a) Typical Crossbar array geometry for various implementations for neuromorphic computing. b) schematic of the sneak-path problem adapted from [155]. Green path is through a high resistance state and red is through three low resistance states.

## VII. APPLICATIONS

In the following we provide a summary of current applications of neuromorphic architectures. These applications are based in crossbar geometries that combine different types of functional memristors (including phase change and metal cation devices). The approach is to emulate all possible existing algorithms in neural networks into arrays of memristor-base cells that allows for vector operations. [163–167]

VII.1 Pattern recognition and data mining

Pattern recognition is at the core of neuromorphic applications using memristor crossbars arrays. For instance, 3x3 matrix crossbar realization have been tested using $TiO_{2-x}$ memristive elements, and up to a 12x12 matrix has also been proposed,[161] see figure 22b. The 3x3 implementation demonstrated a fully operational neural network, based on an integrated, transistor-free crossbar with $TiO_{2-x}$ memristor where the variability was reduced by using a $Al_2O_3/TiO_{2-x}$ heterostructure. In this approach, a single-layer neural network was implemented with ten inputs and three outputs, fully connected with 10x3 = 30 synaptic weights. The goal was to detect a pattern image composed of a 3x3 pixel matrix with black and white elements reproducing the letters "z", "v" and "n" (figure 22c). Noisy datasets for training were fabricated by flipping a pixel randomly and used as training and testing (see figure 22a). High fidelity convergence of network outputs during the training was achieved after six iterations from different initial states as shown in figure 22d.

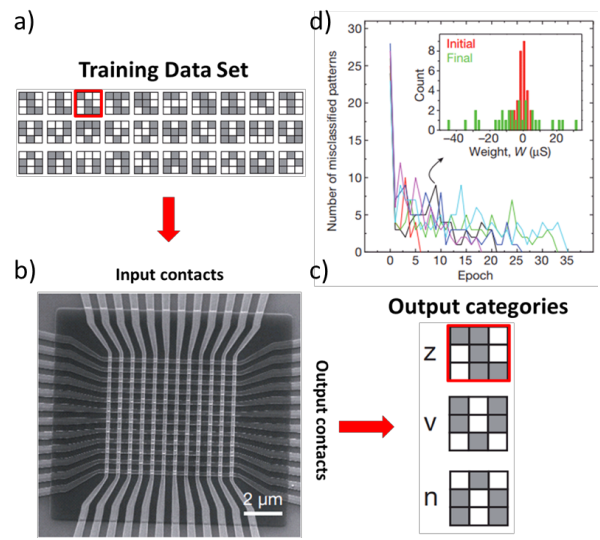

Figure 22. Implementation of pattern recognition using a memristor crossbar array. a) Initial data set used for training the network. b) Device used to implement the crossbar architecture. c) three output categories in which the input data was categorized and d) Performance of the neural network as a function of the number of training sets. Adapted from [161].

Memristors-based crossbar arrays were used to perform matrix operations in an array of nine inputs and two outputs. The inputs were fed with a dataset corresponding to a breast cancer study and the goal was to search for principal components in the data sets. This was implemented using a memristor-based neuromorphic chip using $TaO_x/Ta_2O_5$:Si heterostructure, where the Si doping is used to tailor filament formation. It controls the ion-hopping distance and drift velocity, thus allowing improved tuning of the RS process at the atomic level.

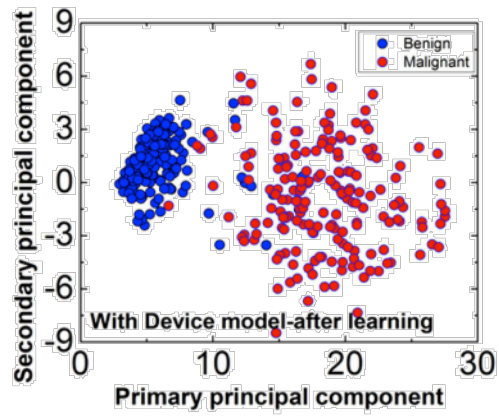

Figure 23. Results of the principal component analysis implemented in a in TaO$_x$/Ta$_2$O$_5$:Si memristor crossbar geometry showing an almost-perfect clustering of a breast-cancer dataset.[155,168]

This work shows that unsupervised learning rules can be successfully implemented in actual devices as depicted in figure 23. The memristor based results are almost identical to standard covariance matrix approaches. This opens new roads for further implementations of neuromorphic computing chips.

VII.2 Analogue computing and other implementations

Memristor-based crossbar architectures have been proposed and recently used to perform sparse coding and image processing [169,170]. Sparse coding representation is a learning method that aims to find a sparse representation of an input data set and therefore constructing a dictionary that allows for identification of patterns in new datasets. The basic elements are not orthogonal, meaning that they may contain similar information and therefore are oversampled. This is developed to emulate the functionality of biological systems which provides the means to analyze complex datasets from input neurons. Some applications of sparse coding are compressed sensing and signal recovery. [171,172]

An implementation of this, using WO$_x$-based analog memristors is shown in figure 24(a). A 32x32 crossbar array was used to implement a sparse coding algorithm [169]. The operational principle consists of finding the basic representation of an element as schematically represented in figure 24(b). The main task is using an image input to decompose it and represent it with a minimal number of "dictionary" elements. The experimental realization of such algorithm is depicted in figure 24(c-d). In figure 24(c) the input image is a 5x5 input matrix where the color code correspond to different conductance states (weighs) that are represented in greyscale. There are 20 dictionary elements, each containing 25 weighs. The reconstructed image is shown in figure 24(c) after the stabilization of the sparse-coding network is reached (figure 24(d)). More complex images were reconstructed proving the concept of sparse coding implementation using memristor-based crossbar architectures.

Other recent implementations of crossbar arrays based on hafnium oxide [170] are as big as 128x64 cells. The implementations described above relays on feature extraction from real-space datasets, such as images. Another important application is the processing of temporal information. In this case, the algorithms are based in a neural network-based computing paradigm called reservoir computing. These type of systems have been recently implemented with WO$_x$ memristors having short-term memory that allows to implement the dynamic reservoir. Applications of such systems have been demonstrated for hand-written digit recognition. [173]

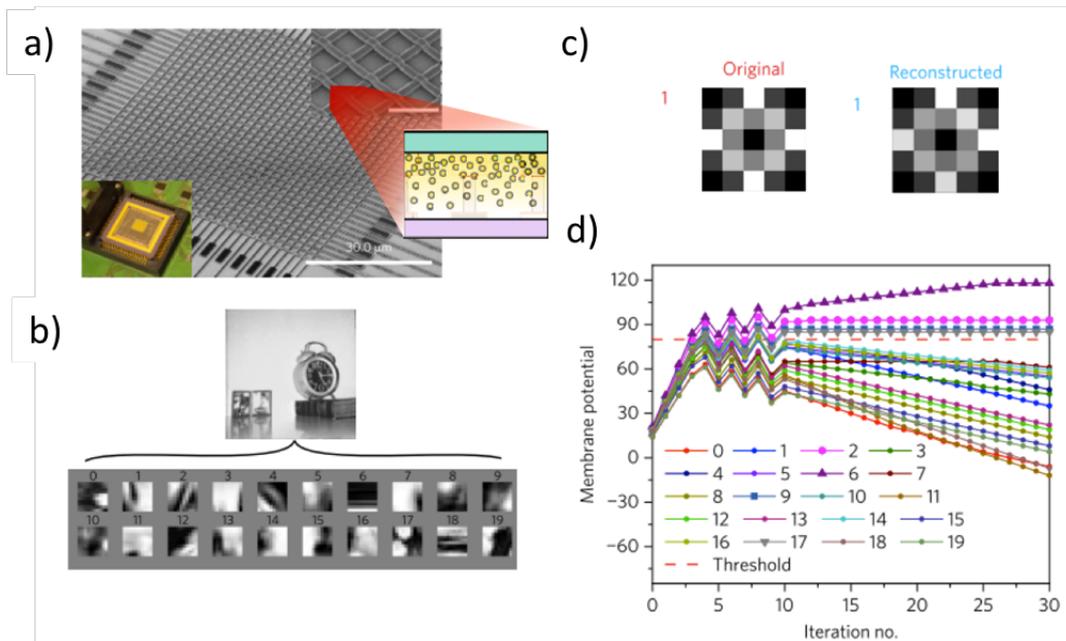

Figure 24. a) A 32x32 crossbar array of WO$_x$-based analog memristors. Each conductance state is controlled by ion redistribution, b) In sparse coding, a dictionary of oversampled elements is constructed which allow for identification of elements in a dataset. As an example the image of a clock is used. c) Original and reconstructed image used to prove the sparse coding concept. d) Membrane potential (voltage output of a neuron) of 20 neurons after each iteration. The dashed line indicated the threshold. After 10 iterations the neurons reach equilibrium. Adapted from [169].

## VIII. CONCLUSIONS AND PERSPECTIVE

We conclude this tutorial discussing some open questions and issues related to the physical implementation of RS for neuromorphic systems. Of course, there are also many challenges related to software implementation, which are beyond the scope of this article. It is important to emphasize that energy-efficient Neuromorphic Computational systems have already been implemented for some limited applications. Thus, there is a clear proof of concept, which shows that this approach should be able to produce a viable computational machine. However, there are several basic research issues in each constituent of such a system, which imply that much research is needed before a general purpose, energy-efficient, computational machine which rivals the human brain can be implemented.

## VIII.1 Immediate Challenges

Immediate development of resistive memories faces two major problems: variability and integration with current technological platforms.

Current RS devices show a great degree of variability, both between different devices and different cycles [174–177]. The fundamental reason is their small size, in the nanometer scale, comparable to important structural lengths such as the grain size or the domain length in Mott insulators. NVRS is highly dependent on defects as they enhance vacancy formation and ion mobility, facilitating the filament formation. Similarly, imperfections in the film act as nucleation centers, reducing the threshold voltage needed to trigger the MIT. Slightly different configurations in each device will result in a great variability between them.

The cycle-to-cycle variability is caused by the irreproducibility of the RS process in the same device, i.e. the device will not switch at the same voltages, or will not have the same resistance every time it is cycled [175–177]. That is also caused by the small sizes involved: filaments tend to grow in an extremely thin fashion, typically a few nanometers across. Set and Reset processes commonly take place in a destructive fashion: part of the filament melts and disconnects and it is very unlikely that the exact same configuration will be repeated when the filament is reformed. Among the TMO, $HfO_X$ and $TaO_X$ have been demonstrated to be particularly reliable, showing less variability between cycles [65,170,177,178].

It is unclear to what extent this device-to-device and cycle-to-cycle variability can affect the performance of RS for Neuromorphic Computing. Some features of biological neural systems are fault tolerant and do not require extreme precision to operate precisely. Querlioz et al. [179] showed that device variations can be overcome by adjusting the neuron firing rate. On the other hand, functionalities such as spike firing are very repeatable between different cells. Whether this variability poses a great problem to real implementations is still not clear.

Short term applications of RS will certainly be integrated together with existing, mature technologies such as CMOS. Nanofabrication procedures might not be compatible with existing technologies might be a complicated task, as some materials might not be suited for that. NVRS commonly uses two electrodes of different metals: one active and one inert, generally Pt or Pd. These materials are incompatible with CMOS technology, so other electrodes, such as TiN or Ru must be used [180]. For VRS, the challenge is even bigger: materials such as $NbO_2$, $VO_2$ or $V_2O_3$ are grown at high temperature (as high as 700 C) in order to display useful MIT properties[181]. Such high temperatures produce deleterious dopant redistribution in the CMOS transistors. New fabrication routes for Mott insulators close to room temperature are needed to overcome this problem.

## VIII.2 Long term challenges

Apart from these immediate challenges, long term implementation of neuromorphic devices will require to address other issues. A selection of these are listed below

General:

1. Should current Neuromorphic software paradigms guide the development of Neuromorphic hardware?
2. Can a general purpose, realistic, energy efficient Neuromorphic machine be designed?
3. Which is the most promising approach for the implementation of Neuromorphic functionalities: charge, spin-torque, superconducting or hybrid systems?
4. Can the charge-based approach eventually scale to produce a computational machine with sufficient elements, flexibility and functionality to rival the human brain?

Materials:

1. Can different neuromorphic functions be implemented in the same or similar compatible materials systems?
2. Are strongly correlated systems the best candidates to search for non-linear phenomena and novel properties useful for the implementation of neuromorphic functionalities?
3. Within the charge-based systems in which materials is it possible to integrate the widest range of relevant functionalities?
4. Which materials systems are the most promising for the implementation of resistive switching, volatile and non-volatile?
5. What is the role of disorder and how fault tolerant should a system be?
6. What is the role of disorder and heat dissipation as a system is downscaled?

Phenomena and devices:

1. Which is the most useful switching mechanism for incorporation into a neuromorphic system?
2. Can ionic motion be controlled precisely enough to produce statistically reproducible results after many cycles?
3. What is the role of thermodynamics at the small scale?
4. How fast can the dynamics of switching be and is this an important issue?
5. Can pure electric switching be implemented or a thermal component will always be present?
6. Can thermal switching be more practical and easier to control than electric switching?
7. Can direct imaging of changes in neuromorphic devices be implemented and studied in-operando in real time?
8. Is there a minimum number of interconnected devices that produce emergent functionalities, which do not arise from the simple sum of individual components?

Architecture

1. What is the role of architecture?
2. Are there any clever artificial architectures which can address and/or illuminate important issues related to interconnectivity?
3. How should the different components of a neuromorphic system be interconnected?
4. Can 3D stacks of crossbar arrays be implemented?
5. Will thermal dissipation be an issue?

This tutorial described the basis and important issues which arise towards the implementation of energy efficient, general-purpose, neuromorphic computer. The paper shows evidence that Resistive Switching is a feasible mechanism and a serious competitor towards the implementation of a neuromorphic machine. The examples described above illustrates that neuromorphic functionalities using Resistive Switching have already been developed although for very limited applications. The main aim and ultimate goal of scaling and incorporating sufficient number of devices into an energy efficient neuromorphic machine which rivals the brain remains an open problem.

**ACKNOWLEDGEMENTS**

This multidisciplinary review paper integrates and includes major components in the fields of; quantum materials, bio-inspired electronics and neuromorphic computation. The development of bio-inspired hybrids (IKS) was supported by the Vannevar Bush Faculty Fellowship program sponsored by the Basic Research Office of the Assistant Secretary of Defense for Research and Engineering and funded by the Office of Naval Research through grant N00014-15-1-2848. The international collaboration between UCSD and CNRS (IKS, MJR) and a major effort to develop an energy-efficient neuromorphic computer is funded through an Energy Frontier Research Center funded by the U.S. Department of Energy, Office of Science, Basic Energy Sciences under Award # DE-SC0019273.

MJR acknowledges fruitful collaborations and discussions with C. Acha, L. Cario, B. Corraze, I.H. Inoue, E. Janod, P. Levy, M.J. Sanchez, P. Stoliar and F. Tesler, and the support from the LIA CNRS-UCSD. JGR acknowledge support from FAPA program through Facultad de Ciencias and Vicerrectoria de Investigaciones of Universidad de los Andes, Bogotá Colombia and Colciencias No. 120471250659 and No. 120424054303. JdV thanks Fundación Ramón Areces for their funding.